\documentclass[letterpaper,12pt]{article}   
\usepackage{osajnl2} 
\usepackage[draft]{hyperref} 

\usepackage{units}
\usepackage{subfig}
\usepackage{amsmath}
\usepackage{amssymb}
\def \D{{\rm d}}

\begin{document}

\title{Measuring small absorptions exploiting photo-thermal self-phase modulation}


\author{Nico Lastzka$^1$, Jessica Steinlechner$^1$, Sebastian Steinlechner$^1$
and Roman Schnabel$^{1,*}$}

\address{$^1$Institut f\"ur Gravitationsphysik,
Leibniz Universit\"at Hannover and Max-Planck-Institut f\"ur
Gravitationsphysik (Albert-Einstein-Institut), Callinstr. 38,
30167 Hannover, Germany}
\address{$^*$Corresponding author: roman.schnabel@aei.mpg.de}

\begin{abstract}
We present a method for the measurement of small optical absorption
coefficients. The method exploits the deformation of cavity Airy peaks that
occur if the cavity contains an absorbing material with a non-zero
thermo-refractive  coefficient $\D n/\D T$ or a non-zero expansion coefficient
$a_{\text{th}}$. Light absorption leads to a local temperature change and to an
intensity-dependent phase shift, i.e. to a photo-thermal self-phase modulation. The absorption coefficient is derived from a comparison of time-resolved measurements with a numerical time-domain simulation applying a Markov-chain Monte-Carlo (MCMC) algorithm.
We apply our method to the absorption coefficient of lithium niobate (LN) doped
with 7\,mol\% magnesium oxide (MgO) and derive a value of $\alpha_\text{LN} =
(5.9\pm 0.9)\times 10^{-4}/{\rm cm}$. Our method should also apply to materials with
much lower absorption coefficients. Based on our modelling we estimate that, with cavity finesse values of the order 10$^4$, absorption coefficients of as low as 10$^{-8}/{\rm cm}$ can be measured. 
\end{abstract}


\maketitle 

\section{Introduction}

Materials with low optical absorption coefficients are essential for
high-precision laser-interferometric measurements. Absorptions in mirror
substrates of as low as $10^{-6}/\rm{cm}$ already limit gravitational wave
detectors because absorption leads to heating and a thermal deformation of the
mirrors \cite{win91} and also to photo-thermal noise \cite{BGV99}. Future
gravitational wave detectors will use cryogenically cooled mirrors \cite{Arai09}
to reduce thermally excited motions of mirror surfaces. Then, low optical
absorptions will become even more crucial. Consequently,
the measurement of small absorption coefficients in the regime below
$10^{-6}/\rm{cm}$ is
important to find appropriate mirror materials and to enable the reliable design
of future gravitational wave detectors, such as the Einstein Telescope
\cite{et-paper,et-gw}. 

In the past, several methods have been developed that are able to measure
absorption coefficients of the order of $10^{-6}/\rm{cm}$. All these methods are based on
indirect measurement schemes. They do not directly sense the power loss of a
transmitted beam but utilize the temperature increase that arises due to the
absorption. In calorimetric approaches the temperature increase is directly
measured \cite{lzh96}. Other approaches exploit light beam deflection or beam
shape deformation due to local heating \cite{lb03,hil06}. 

In this paper we present another indirect measurement scheme to determine small
absorptions. The material under investigation is put inside an optical cavity
whose length is linearly scanned over a cavity Airy peak. Approaching cavity
resonance the temperature along the cavity mode increases and the optical path
length for a cavity round trip changes. The thermally induced optical path
length change is a photo-thermal self-phase modulation resulting in a deformed
shape of the Airy peak. Since the phase change depends on the light intensity it
may be considered as the result of a ``thermo-optic Kerr-effect''. Importantly,
the Airy peak deformation depends on the scan direction, i.e. whether the cavity
is shortened or lengthened. The hysteresis in the time-resolved measurements
provides information of the absorption coefficient, if relevant material
parameters are known and included in a numerical time-domain simulation. A
positive side-effect of our method is the power build-up inside the cavity which
compensates the need for laser sources with higher powers when approaching the
regime of extremely low absorption.

\section {Theory and Method}
\begin{figure}[t]
  \begin{center}
    \includegraphics[width=8cm]{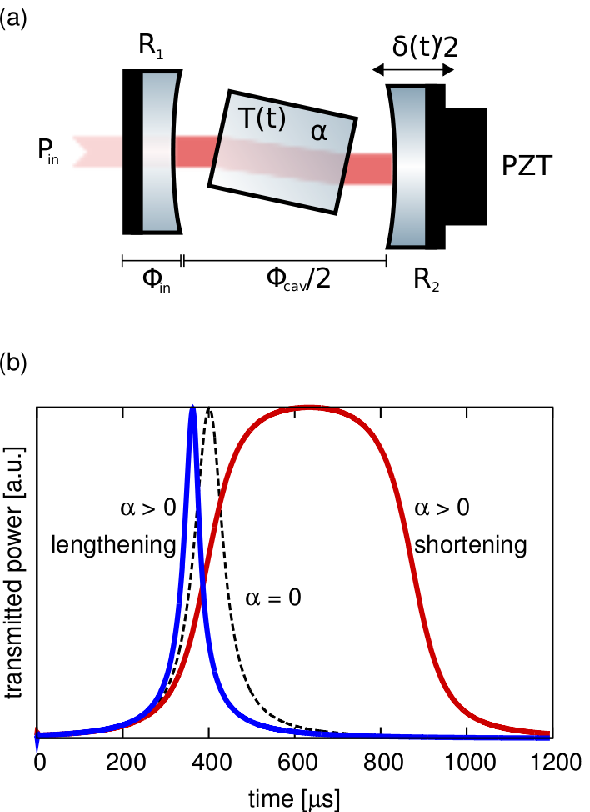}
    \caption{Scheme of the measurement method.
    (a) When changing the cavity round trip phase by $\delta (t)$, absorption of
    intracavity power leads to a temperature change $T(t)$ and an additional
    phase for the intracavity field $\phi_\text{cav}$. $R_1$ and $R_2$ are the
    mirror power reflectivities;   $\phi_\text{in}$ is the input phase. By using
    the Brewster angle, surface reflections can be avoided. 
    (b) Airy peaks showing hysteresis due to the photo-thermal
    self-phase modulation. The
    dashed line shows the peak without any absorption ($\alpha = 0$), whereas the
    blue (narrow) and red (broad) peaks show
    the Airy peak for the same absorption coefficient $\alpha > 0$, but for lengthening and shortening the
    cavity, respectively.}
    \label{fig:heating scheme}
  \end{center}
\end{figure}

In this section we describe the time-domain simulation that is used to analyse
the measurement data and to deduce the absorption from it. Our approach is based
on work by Hello and Vinet \cite{hv90, hv93} in which they describe the heating
of an absorbing material due to a Gaussian laser beam. In our case, a sample of
the absorbing material with polished (plane) surfaces is placed inside a (high)
finesse cavity. {One may choose} the light's angle of incidence to be the
Brewster angle to avoid reflection losses. A schematic is shown in
Fig.~\ref{fig:heating scheme}. When the cavity round-trip phase
$\phi_{\rm{cav}}$ is linearly increased (or reduced) by $\delta (t)$ and
scanned over a cavity resonance, absorption leads to a dynamic temperature
profile inside the material and inside the cavity mirror surfaces. The result is
a (photo-thermal) self-phase modulation and a deformation of the cavity Airy
peak. 

Our time-domain model iteratively calculates the intracavity intensity after
each round trip. The time $t$ is discretized becoming an integer multiple of the
round trip time, yielding $t = t_n = n/\Delta f_\text{FSR}$, where $\Delta
f_\text{FSR}$ is the cavity free-spectral-range. 
The intra-cavity field $a_n := a(t_n)$ after $n$ round trips reads
\begin{equation}
  a_n = \text{i} \sqrt{1 - r_1^2}\, a_\text{in} \text
  {e}^{\text{i}\phi_\text{in}(t_n)} + r_1
  \tilde{r}_2\, \text{e}^{\text{i}\phi_n(t_n)} a_{n-1}\,.
\end{equation}
Here, $r_1$ is the amplitude reflectivity of the first mirror, whereas $\tilde{r}_2$ is the effective
amplitude reflectivity of the second mirror, which includes all round trip
losses. The amplitude of the incident power $P_\text{in}$ is given by $a_\text{in} =
2\sqrt{P_\text{in}/(\epsilon_0 c\, \pi w_0^2)}$, where $w_0$ is the waist radius
of the beam $\epsilon_0$ the dielectric constant and $c$ the speed of
light. 
The cavity input field gains the phase $\phi_{in}$ that is due to the
temperature gradient inside the incoupling mirror emerging from its coating
absorption.
The phase $\phi_n(t_n)$ after $n$ round trips can be written as
$$\phi_n=\delta (t_n)+\phi_\text{spm} (t_n, \alpha)\,,$$
where $\delta (t_n)$ is the phase due to the external cavity detuning and 
$\phi_\text{spm} (t_n, \alpha)$ is due to the photo-thermal (internal)
self-phase-modulation which depends on the absorption $\alpha$.
The external detuning for the round trip number $n$ is determined from
\begin{equation}
  \delta (t_n) = \delta_0 + n 2 \pi N_\text{FSR}\frac{\omega_s}{\Delta
  f_\text{FSR}}\,.
\end{equation}
Here, $N_\text{FSR}$ is the number of free spectral ranges that were scanned  with
frequency $\omega_s$. The velocity $v_\text{m}$ of the scanning mirror is therefore given by
\begin{equation}
  v_\text{m} = 2\lambda\omega_s\cdot N_\text{FSR} \,.
\end{equation}

The temperature distribution $T_n$ for round trip $n$ is calculated by
using the recurrence relations (Eq. (15) in \cite{hv93}). These equations
determine the radial and longitudinal temperature gradient at any time including
thermal conductivity. The starting point is
the external temperature $T_0$. The detuning $\phi_\text{spm}$ induced by the
photo-thermal self-phase-modulation is then given by equations (33) and (35) of
\cite{hv90}. Note that $\phi_\text{spm}$ includes the effects of a non-zero
thermo-optic coefficient $\D n/\D T$ and a non-zero expansion coefficient
$a_\text{{th}}$.
As a starting point for the numerical simulation we use the steady-state
solution for the start detuning
$\delta_0$
\begin{equation}
  a_0 = \text{i} \sqrt{1 - r_1^2}\, \frac{a_\text{in}}{1 - r_1 \tilde{r}_2 e^{\text{i} \delta_0}}\,.
\end{equation}
Assuming a perfect mode-matching of the input field $a_\text{in}$ to the cavity mode, the reflected and transmitted fields are given by
\begin{align}
\label{a_refl}
  \begin{split}
    a_\text{refl}(t_n) & = \text{i} \sqrt{1 - r_1^2}\,\, a_n(\alpha) + r_1 a_\text{in}
    \text{e}^{\text{i}\phi_\text{in}} \, ,\\
    a_\text{trans}(t_n) & = \text{i} \sqrt{1 - \tilde{r_2}^2}\, a_n(\alpha)\,.
  \end{split}
\end{align}

Eq.~(\ref{a_refl}) can now be used to calculate the time-resolved shape of an
Airy peak. By varying $\alpha$ the result can be fitted to the measurement
performed in reflection or transmission of the cavity. Figure \ref{fig:heating
scheme}(b) shows an example, i.e. simulated Airy peaks obtained from a cavity
containing some absorbing material with $\D n/\D T > 0$. The resonance peaks get
broader when shortening the cavity because the positive thermo-refractive
coefficient counteracts the external change of the cavity length. Accordingly
the resonance peaks get narrower when the cavity length is increased.
In particular the hysteresis can be used to precisely determine the absorption
of the material.  For comparison we also show the normal Airy peak without
self-phase modulation (dashed line).

\section{Measurements and data analysis}
To characterize the feasibility of our method we performed a series of
absorption measurements on a $\unit[7]{mol \%}$ MgO-doped LiNbO$_3$ crystal. A
single measurement {set} involves a characterization of the piezo electric
element that is used to change the cavity length, and altogether four
time-resolved photo-electric detections. A fast photo-diode records the Airy
peaks in reflection of the cavity when the latter is (a) lengthened or (b)
shortened, using (1) a low laser power without any thermal Airy peak deformation
or (2) a laser power at which a thermal deformation is clearly visible. The
low-power setting is used to quantify the two reflectivities $r^2_1$ and
$\tilde{r}^2_2$. The high-power setting is used to quantify the absorption
coefficient $\alpha_{\rm LN}$. All three quantities and their error bars are
deduced from a \textit{single} measurement {set} and a numerical time-domain
simulation applying a {Markov}-chain Monte-Carlo (MCMC) algorithm. Records (a1)
and (b1) are identical thereby confirming that the laser power was chosen to be
low enough {so that thermal effects do not yet come into play}. Note that (a1)
and (b1) will not necessarily have the shape of the central Airy peak (dashed)
shown in Fig.~\ref{fig:heating scheme}b but may show a ringing effect due to
the cavity loading or decay time \cite{MBBCHM00}. This effect is also precisely
modelled in our simulation.

\subsection{Experimental setup}

In our research group, we routinely use cavities containing MgO-doped
LiNbO$_3$\;(LN)\;crystals for second harmonic generation (SHG) and squeezed
light generation (SLG) at a wavelength of 1064\,nm
\cite{CVDS07,Vahlbruch08,vah08}.
The optical absorption of these nonlinear crystals is a limiting factor in
achieving high conversion efficiencies and high squeezing factors. Accurate
absorption coefficients are therefore required to optimize the nonlinear cavity
design. Unfortunately, manufacturers' data typically are rather inaccurate and a
standard value of $\alpha_\text{LN}\lesssim 10^{-3}/\rm{cm}$ at 1064\,nm is
quoted in most cases. In this work we used one of our SHG cavities to measure
the absorption coefficient of LN and test our new absorption measurement
technique.

\begin{figure}[t]
  \begin{center}
    \includegraphics[width=7cm]{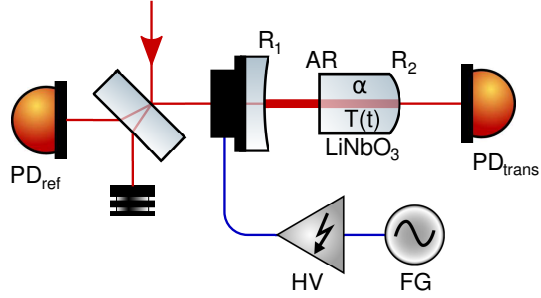}
    \caption{Experimental setup: The resonator is formed by the incoupling
    mirror with reflectivity $R_1 \approx 90\,\%$ and the crystal's highly reflecting
    (HR) coating with a reflectivity of $R_2 > 99.8\,\%$. The resonator length
    is scanned with a frequency from a function generator (FG) which is fed
    through a high voltage amplifier (HV).}
    \label{fig:experimentalsetup}
  \end{center}
\end{figure}

Figure \ref{fig:experimentalsetup} shows the experimental setup. An incoupling mirror and the curved side of the plano-convex LiNbO$_3$ crystal form a single-ended standing wave cavity for laser light at 1064\,nm. The cavity mirrors have power reflectivities of $r^2_1=R_1 \approx 90\,\%$ and $r^2_2=R_2 > \unit[99.8]\%$, respectively. A small air gap separates the incoupling mirror from the anti-reflection coated, plane crystal surface. Table~\ref{tab:parameters} contains {detailed} geometric parameters of this resonator and the laser beam as well as the material parameters of the LiNbO$_3$-crystal. 

Up to 1.5\,W of single mode radiation at 1064\,nm was modematched into the
cavity with a modematching efficiency of greater than 95\%. To prevent the generation
of second harmonic radiation, both the input field polarization and the crystal
temperature were detuned from their usual operation point. A piezoelectric
transducer (PZT) moved the incoupling mirror to allow for a scan of the cavity
length. {The photo diode measured} the temporal behaviour of the reflected laser
power. {We ensured that the photodiode was fast enough, i.e. had a high
bandwidth, so that it did not influence the shape of the recorded Airy peaks.}



\begin{figure}[t]
  \begin{center}
    \includegraphics[height=6cm]{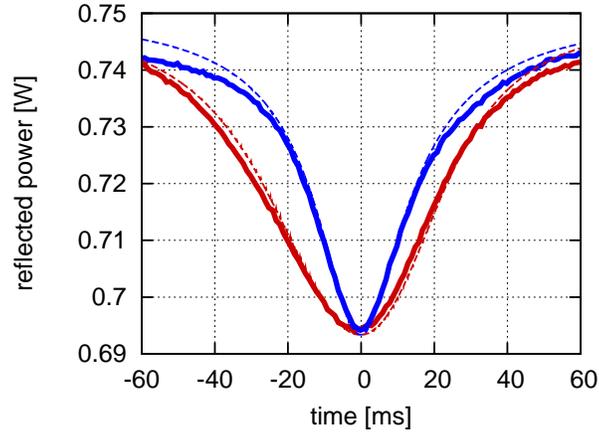}
	   \caption{Example of measured (solid) and simulated (dashed) Airy peaks with visible thermal effect.
  Without absorption all curves would be identical. The curves were measured in reflection, no parameter
           other than the scan direction was changed. The red curve forms for a
           shortening resonator, the blue one for a lengthening resonator. }
    \label{fig:peaks}
  \end{center}
\end{figure}

Fig. \ref{fig:peaks} shows an example of Airy peaks with visible thermal effects
as measured in reflection of the cavity. The blue curve forms for a lengthening
resonator, the red curve for a shortening resonator. No parameter other than the
scan direction was changed. The two curves would be identical without self-phase
modulation and no hysteresis effect would occur without absorption. The solid
lines in Fig. \ref{fig:peaks} represent our simulation fitted to the
experimental data. The narrow curves show a discrepancy in the left wings, the
broad curves show a discrepancy in the right wings. The two deviations come from the
non-perfect modematching to the cavity and the excitation of a higher-order
cavity mode.


{Apart from taking a simple full measurement set, i.e. lengthened and shortened
resonator at two different laser powers, we performed measurements at three
different laser powers and three different scan frequencies. While not strictly
necessary for an absorption measurement, these measurements demonstrate the
consistency of our result, see below.}

\subsection{Measurement analysis}
For the analysis of the measured peaks the PZT had to be calibrated because of
its own hysteresis and non-linearity.  This calibration was done at low laser
powers where no thermal effect occurred. We measured the width of the Airy peaks at
different positions of the PZT's scanning range by slightly shifting the laser
frequency. A third-degree polynomial well described the peak width depending on
peak position. Together with a scan showing a full free spectral range, we used
this polynomial to linearize the PZT movement. We performed the calibration for
both scan directions and for each scan velocity that we used.  Measurements
were performed at three different scan velocities, namely
$2\cdot\unit[1064]{nm}/\unit[5]{ms}$, $2\cdot\unit[1064] {nm}/\unit[2.5]{ms}$
and $2\cdot\unit[1064]{nm}/\unit[0.285]{ms}$.  For each scan velocity we measured
Airy peaks at three different input powers: $\unit[100]{mW}$, $\unit[750]{mW}$
and $\unit[1.5]{W}$.

Table~\ref{tab:parameters} gives a complete list of the parameters that enter
our simulation. We used values from literature for the material parameters, for
the geometric parameters we chose values to our best knowledge of the cavity
design. The mirror reflectivities $R_1$ and $\tilde{R}_2$ define the cavity resonance
width and power build-up. Here, $\tilde{R}_2$ is an effective reflectivity,
which includes absorption and scattering losses. Only this value, rather than
the pure reflectivity $R_2$, is accessible when light enters the cavity through
mirror $R_1$. As resonance width and power build-up have a strong impact on the heating of the
substrate, we do not use the reflectivity values as given by the
coating manufacturer. Instead, we treat $R_1$ and $\tilde{R}_2$, as well as the
absorption $\alpha_{\rm LN}$, as free parameters of our simulation.

For low input powers and fast scan velocities the resulting temperature change
inside the substrate is small and no deformation of the peaks is visible.
Such time series are optimally suited to determine $R_1$ and $\tilde{R}_2$.
Towards higher input powers
and lower scan velocities, the peaks begin to show a hysteresis. For all our
measurements that were performed with different laser powers, the hysteresis
values
could be explained completely by the self-phase modulation, i.e. our simulation
provided a very good discription of the measurement. From this we conclude that
no spatial deformation of the cavity mode occurred. 

\begin{figure}[t]
\centering
	\hspace{-2mm}{\includegraphics[height=5.8cm]{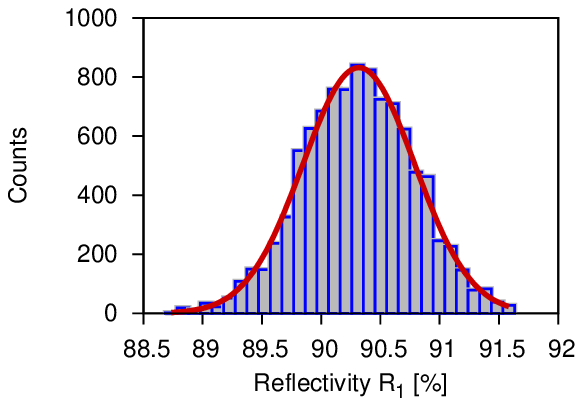}}\\
	{\includegraphics[height=5.8cm]{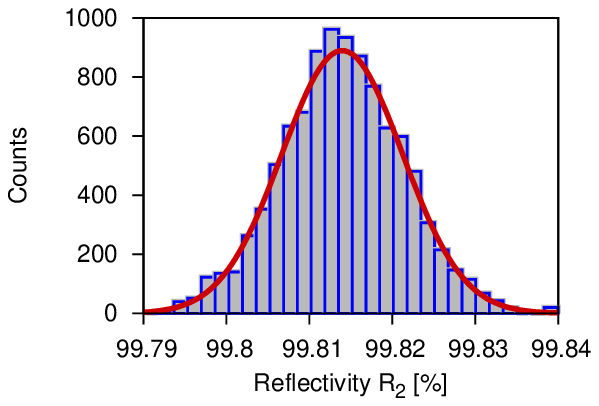}}\\
  {\includegraphics[height=5.8cm]{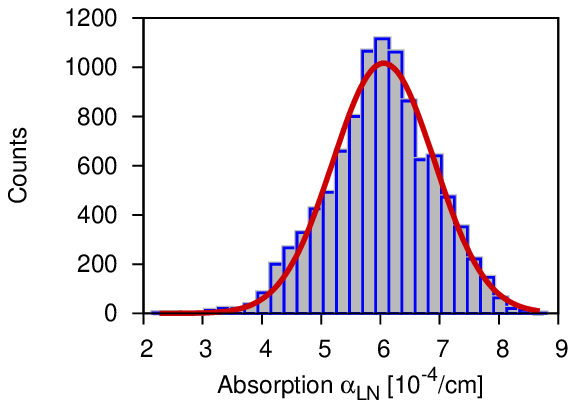}}
        \caption{The Metropolis-Hastings MCMC algorithm draws samples from the
        simulation parameter space creating a chain of individual realizations
        that result in the parameter distributions. Here we show the histograms
        of a chain obtained from
        a single measurement set at a laser input power of $\unit[0.75]{W}$ and
        a scan-velocity of $v= 2\cdot \unit[1550]{nm}/\unit[2.5]{ms}$. $R_1$ (top) and
        $\tilde{R}_2$ (middle) are required to characterize the cavity. The
        bottom figure shows the result for $\alpha_{\rm LN}$. The
        bars represent histograms of the MCMC run. The curves are
        gaussian fits to the histogramms.}
\label{fig:histogram}
\end{figure}

We performed a quantitative analysis by calculating the variance between
simulated and measured data.  Starting from an initial set of parameters, we
ran a Metropolis-Hastings Markov-chain Monte Carlo (MCMC) \cite{mcmc} algorithm which
minimized the variance. The data chains generated can be
converted into histograms for the free simulation parameters. The histograms
for the reflectivities $R_1$, $\tilde{R}_2$ and for the absorption $\alpha_{\rm
LN}$ as derived from a single measurement setting are shown in Figure
\ref{fig:histogram}. As the histograms closely resemble Gaussian distributions,
we give the mean value and standard deviation of all nine measurements in
Table~\ref{tab:results}.

\begin{figure}[t]
  \begin{center}
    \includegraphics[height=7cm]{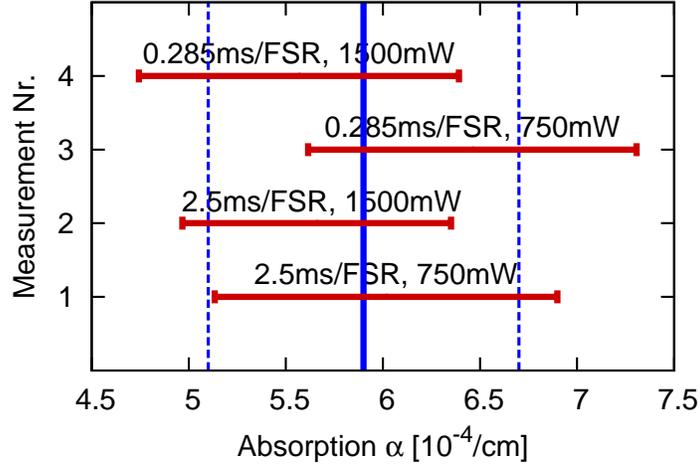}
	   \caption{Four independent measurement values of the absorption coefficient $\alpha_{\rm LN}$ and their statistical standard deviations. The blue lines show the mean value of the four measurements,
           which is $\alpha_\text{LN} = 5.9\times 10^{-4}/\rm{cm}$. The dashed blue lines
           mark the averaged standard deviation of $\Delta\alpha_\text{LN} =
           0.8\times 10^{-4}/\rm{cm}$.}
    \label{fig:alphaerrorbars}
  \end{center}
\end{figure}

The mean value of the results for the incoupling mirror was found to be $R_{1} =
(89.43\pm 0.75)\,\%$ which is in good agreement with the manufacturer's
specification for this coating ($(90\pm 1)\,\%$).  The effective value for the
high-reflective coating of the crystal was determined to be $\tilde{R}_2 =
(99.79\pm 0.01)\,\%$, which is also in accordance with the specifications.  The
measurement with $\unit[100]{mW}$ input-power at a scan velocity of $v= 2\cdot
\unit[1064]{nm}/\unit[5]{ms}$ was the boundary where a small thermal effect was
visible. However, no accurate absorption coefficient could be deduced due to
rather large error bars. Four measurements showed a significant thermal effect
and were used to derive four independent values for the absorption coefficient
of LN. All four values for $\alpha_{\rm LN}$ have mutually overlapping error
bars. Figure \ref{fig:alphaerrorbars} gives a graphical overview of the results
for $\alpha_{\rm LN}$ for all measurements. The mean value of the four results
is $\alpha_\text{LN} =5.9\times 10^{-4}/\rm{cm}$. As the error bar we quote the
standard deviation (an averaged value) of the single measurement set which
typically was $\pm 0.8\times 10^{-4}/\rm{cm}$. This number includes the influence from
errors in the reflectivities $R_1$ and $\tilde{R}_2$ as shown in
Fig.~\ref{fig:histogram}. 



\subsection{Error propagation}
We considered the influence of possible errors in the input parameters on the
resulting value for $\alpha_\text{LN}$ (from a single measurement). For this
investigation we individually changed the values of the simulation input
parameters and recalculated $R_1$, $\tilde{R}_2$ and $\alpha_\text{LN}$ for
each case.

Our investigation showed that the parameters can be grouped into two
categories.  The first category contains parameters that have a very weak
influence on the absorption coefficient in our case. For our system, heat
radiation described by the material emissivity $0.0 < \epsilon \leq 1.0$ is not
relevant at all, because the substrate is heated only within the beam radius,
far away from the substrate's surface. Also the absorption coefficient of the
substrate coatings ($\alpha_{\rm coating}$) can be neglected, as it is much
smaller than the substrate absorption $\alpha_{\rm LN}$ and the coating
thickness is negligible compared to the substrate dimension. A few percent
change of the values for the index of refraction $n$, the intra-cavity airgap
$s$, the substrate radius $R$, and the beam waist $\omega_{0}$ also has a
negligible effect on the absorption coefficient $\alpha_{LN}$. The second
category contains the remaining parameters of our model. These parameters and
their respective influence on $\alpha_{LN}$ for a $\unit[4]\%$ change in the
parameter value are the input laser power $P$ ($\unit[3.6]\%$), the substrate
length $L$ ($\unit[3.6]\%$), the thermal conductivity $k_\text{th}$
($\unit[1.6]\%$), the thermal refractive coefficient $\D n/\D T$ ($\unit[3.6]\%$), the
thermal expansion $a_\text{th}$ ($\unit[1.4]\%$), the density $\rho$ and the
heat capacity $c$ ($\unit[2.8]\%$). Note that in the simulation $\rho$ and $c$
always appear as a product, and the influence of their error bars is identical.
 
Assuming that our measured parameters as well as the material parameters from
literature are precise to within \unit[4]{\%} and statistically independent
from each other, we conclude that the error of $\pm 0.8\times 10^{-4}/\rm{cm}$
($\pm\unit[13.6]{\%}$) coming directly out of the Markov-chain Monte-Carlo
simulation dominates the error on our final result. The total error sums up to
$\pm 0.9\times 10^{-4}/\rm{cm}$ ($\pm\unit[15.7]{\%}$).

\subsection{Sensitivity of the method}
\begin{figure}[t]
 \begin{center}
  \includegraphics[height=14cm]{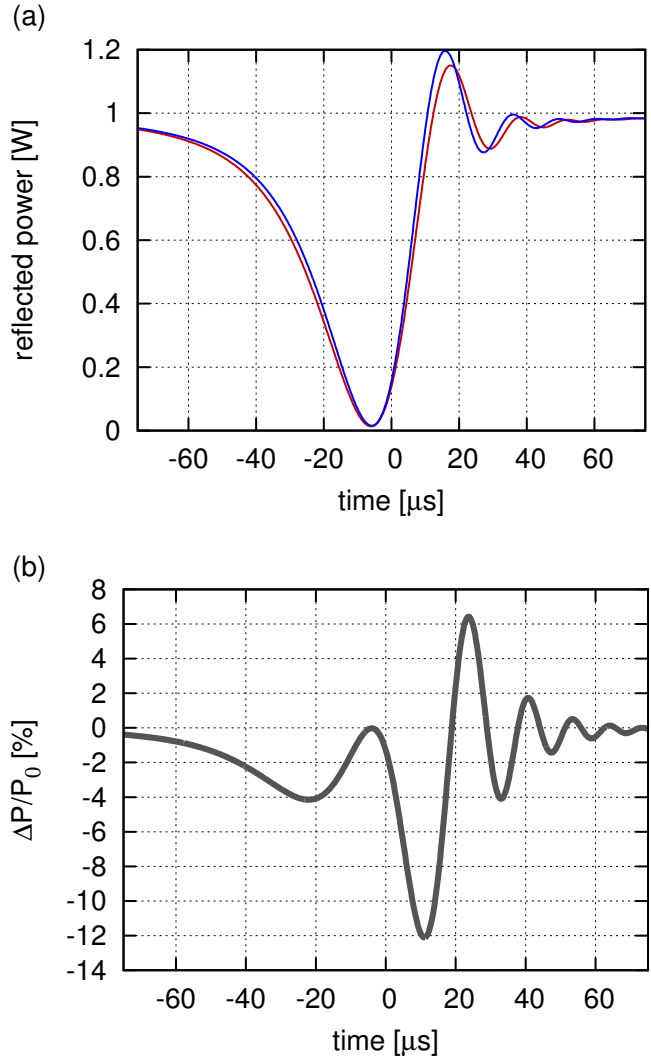}
 \end{center}
 \caption{Simulated hysteresis effect for the Airy peaks in reflection from a
 monolithic silicon cavity of finesse 20000. The curves are normalized to the
 input power of 1 W at 1550 nm. The scan velocity of the cavity length is
 $2\cdot 1550$ nm/s and the absorption was assumed to be $10^{-8}/\rm{cm}$. Other
 parameters can be found in Table~\ref{tab:parameters}. (a)
 Airy peaks for lengthening (blue) and shortening (red) the cavity. (b) The
 difference of the two scan directions $\Delta P$ normalized to the incident laser
 power of 1 W.}
  \label{fig:accuracy estimation}
\end{figure}

To make a prediction of the sensitivity of our method, we consider the
absorption measurement of crystalline silicon at a wavelength of 1550 nm. This
value has not been measured before, but data at shorter wavelengths
\cite{GKe95,KG94}
suggest an absorption coefficient smaller than $10^{-8}/\rm{cm}$ in case of pure
silicon \cite{Schnabel10}.
Our simulation is based on a 6.5\,cm long silicon sample inside a cavity of
finesse 20,000 pumped with 1 W of input laser power. The reflectivities of the
two cavity mirrors are assumed to be identical. Fig.~\ref{fig:accuracy
estimation}\,a shows the Airy peaks as detected in the reflected light for both
cavity scan directions with an absorption of $10^{-8}/\rm{cm}$. The scan velocity
used in that simulation was $v_\text{m} = 2\cdot 1550$ nm/s. The curves are normalized
to the input power of 1 W. Both curves show oscillations and values above unity
which arise from the cavity loading and decay time \cite{MBBCHM00}.
Fig.~\ref{fig:accuracy estimation}\,b shows the difference of the two scan
directions normalized to the Airy peak without absorption. We find a significant
hysteresis curve that reaches up to 12\% of the input power. Our simulation
neglects the influence of the absorption in the dielectric coatings. In
practice, the absorption inside the cavity mirror coatings has to be insignificant
as in our experiment or it has to be measured
independently when the sample is removed from the cavity. Anti-reflection
coatings or high-reflection coatings on the sample itself can also be taken into
account when two different sample lengths are studied. Generally, the photo-thermal
self-phase modulation from the coating absorption must not dominate the
overall photo-thermal effect inside the cavity. New low-loss coating materials
such as diamond \cite{SBD81} or monolithic, nano-structured surfaces
\cite{Brueckner10} might be used.   

\section {Conclusion}
In this paper we introduce a new low-absorption measurement method based on the
optical phase change inside the material when absorption leads to local heating.
The effect is understood as a cavity-assisted photo-thermal self-phase
modulation of light. We used our method to determine the absorption coefficient
$\alpha_{\rm LN}$ of a LiNbO$_{3}$ crystal.  Our result of $\alpha_\text{LN} =
(5.9\pm 0.9)\times 10^{-4}/\rm{cm}$ is in accordance with the typically referred upper
bound of $10^{-3}/{\rm cm}$ as available on manufacturer websites. Measurements
with different laser powers could all be well described without considering a
spatial mode distortion. We conclude that no such mode conversion occurred in
our experiments. However, this might be possible at even higher laser powers or
smaller waist sizes. We
theoretically applied our method to a material with an absorption coefficient of
$\alpha = 10^{-8}/\rm{cm}$. We conclude that such low absorptions should be
measureable when a sample of a few cm length is put into a cavity with a finesse
of the order 10$^4$. Our time-resolved Markov-chain Monte-Carlo (MCMC)
simulation is based on a variety of material parameters. The coupling of
parameter errors into the error of the absorption coefficient is linear or
less.\\

\section*{Acknowledgments}
We thank Harald L\"uck for many valuable remarks on the manuscript. This
research was supported by the Centre for Quantum Engineering and Space-Time
Research, QUEST. We also acknowledge funding from the International
Max Planck Research School (IMPRS) on Gravitational Wave Astronomy.

\listoffigures

\pagebreak[4]
\pagebreak[4]
\pagebreak[4]
\pagebreak[4]
\pagebreak[4]
\pagebreak[4]
\begin{table}[t]
 \centering
  \caption{Material and geometric parameters of the LiNbO$_3$- and Si-samples
  and cavity geometric parameters used for the simulations.}
 \label{tab:parameters}
 \begin{tabular}{lll}
 \hline
        Material parameters & LiNbO$_3$ & Si \\
 \hline
 index of refraction $n$	& $2.147$ \cite{zel97} & 3.48 \cite{flm} \\
        thermal refr. coeff. $\D n/\D T$ & $\unit[38.5\cdot10^{-6}/]{K}$
        \cite{jun90} & $\unit[176.0\cdot 10^{-6}]{/K}$\cite{flm} \\
        specific heat $c$ & $\unit[630]{J/(kg\,K)}$ \cite{web} &
        $\unit[713]{J/(kg\,K)}$ \cite{hull}\\
        density $\rho$ & $\unit[4635]{kg/m^{3}}$ \cite{kim95} &
        $\unit[2330]{kg/m^3}$ \cite{web}\\
        thermal expansion $a_{\rm th}$ & $\unit[14.8\cdot10^{-6}/]{K}$
        \cite{web} & $\unit[2.53\cdot 10^{-6}/]{K}$ \cite{hull}\\
        thermal conductivity $k_{\rm th}$ & $\unit[4.19] {W/(m\,K)}$
        \cite{won02} & $\unit[1.56]{W/(m\,K)}$ \cite{gs64}\\
        material emissivity $\epsilon$ & $1.0^{a}$ & $1.0^a$\\
        coating absorption $\alpha_{\rm coating}$ & $\unit[0.0] {/cm}$ &
        $\unit[0.0]{/cm}$\\
        \hline
        Cavity geometric parameters &  & \\
  \hline
        airgap $s$ & $\unit[24]{mm}$ & \unit[0]{mm}\\
        beam waist $\omega_0$ & \unit[24]{$\mu$m} & \unit[160]{$\mu$m}\\
        crystal length $L$ & $\unit[6.5]{mm}$ & \unit[65.0]{mm}\\
        crystal radius $R$ & $\unit[2]{mm}$ & \unit[50.0]{mm}\\
 \hline
 \end{tabular}
 \hspace{0.5 cm}
\begin{flushleft}
 $^{a} 0.0 < \epsilon \leq 1.0$ are the boundaries for the thermal emissivity.
 For our systems the value of this parameter is not relevant since $R \gg
 \omega_0$.
\end{flushleft}
 \end{table}
\pagebreak[4]

\begin{table*}[t]
 \centering
 \vspace{0.5 cm}
 \caption{Results for $R_1$, $\tilde{R}_2$ and $\alpha_{\rm LN}$: Mean values as well as standard deviation of the parameters are given.}
 \label{tab:results}
 \begin{tabular}{l|l|ll|ll|ll}
 \hline
 	$f$ in 			&$P$ in		&\multicolumn{2}{c}{$R_1$}
        &\multicolumn{2}{c}{$\tilde{R}_2$} &\multicolumn{2}{c}{$\alpha_{\rm LN}$
        in $10^{-4}/{\rm cm}$}\\
        ${\rm ms}/\Delta f_{\rm FSR}$	&W		& $\overline{R}_1$
        &$\Delta R_1$ $\cdot10^{3}$	&$\overline{\tilde{R}}_2$
        &$\Delta \tilde{R}_2$	$\cdot10^{5}$	&$\overline{\alpha}_{\rm LN}$
        &$\Delta \alpha_{\rm LN}$\\
 \hline
	0.285			&0.1		&0.89668	&6.46	&0.99812	&8.58		&-				&-\\
	0.285			&0.75		&0.89585	&5.37	&0.99793	&8.3		&-				&-\\
	0.285			&1.5		&0.88532	&3.62	&0.99786	&5.12		&-				&-\\
	2.5				&0.1		&0.8957		&19.2	&0.99802	&26.6		&-				&-\\
	2.5				&0.75		&0.90316	&4.68	&0.99814	&1.27		&6.016		&0.8828\\
	2.5				&1.5		&0.88438	&7.74	&0.9978		&12.2		&5.5685		&0.692\\
	5					&0.1		&0.90153	&5.26	&0.99804	&7.89		&(10.247)	&(4.39)\\
	5					&0.75		&0.89401	&7.41	&0.99797	&11.5		&6.4605 	&0.846\\
	5					&1.5		&0.892		&11.8	&0.99746	&24.1		&5.5672		&0.824\\
	\hline
 \end{tabular}
\end{table*}

\end{document}